\documentclass[sigconf,nonacm]{acmart}
\pdfoutput=1
\hypersetup{keeppdfinfo}

\usepackage{graphicx}
\usepackage{amsmath}
\usepackage{booktabs}
\usepackage{algorithm}
\usepackage{algorithmic}
\usepackage{amsfonts}
\usepackage{multirow}
\usepackage{makecell}
\usepackage{subfigure}
\usepackage{color}
\usepackage{bm}
\usepackage{url}
\usepackage[cal=cm]{mathalfa}
\usepackage{balance}
\usepackage{threeparttable}
\usepackage{lipsum}
\usepackage{enumitem}
 \usepackage{pifont}
 \usepackage{tabularx}
 \usepackage{makecell}
 \usepackage{float}

\AtBeginDocument{%
  \providecommand\BibTeX{{%
    \normalfont B\kern-0.5em{\scshape i\kern-0.25em b}\kern-0.8em\TeX}}}

\begin{document}
\title{CRM: Retrieval Model with Controllable Condition}

\renewcommand{\shorttitle}{CRM}

\author{Chi Liu}
\affiliation{
  \institution{Kuaishou Technology}
  \city{Beijing}
  \country{China}
}
\email{liuchi05@kuaishou.com}

\author{Jiangxia Cao}
\affiliation{
  \institution{Kuaishou Technology}
  \city{Beijing}
  \country{China}
}
\email{caojiangxia@kuaishou.com}

\author{Rui Huang}
\affiliation{
  \institution{Kuaishou Technology}
  \city{Beijing}
  \country{China}
}
\email{huangrui06@kuaishou.com}

\author{Kuo Cai}
\affiliation{
  \institution{Kuaishou Technology}
  \city{Beijing}
  \country{China}
}
\email{caikuo@kuaishou.com}

\author{Weifeng Ding}
\affiliation{
  \institution{Kuaishou Technology}
  \city{Beijing}
  \country{China}
}
\email{dingweifeng@kuaishou.com}

\author{Qiang Luo}
\affiliation{
  \institution{Kuaishou Technology}
  \city{Beijing}
  \country{China}
}
\email{luoqiang@kuaishou.com}

\author{Kun Gai}
\affiliation{
  \institution{Unaffiliated}
  \city{}
  \country{}
}
\email{gai.kun@qq.com}

\author{Guorui Zhou}
\affiliation{
  \institution{Kuaishou Technology}
  \city{Beijing}
  \country{China}
}
\email{zhouguorui@kuaishou.com}

\begin{abstract}
Retrieval models for recommendation need to account for outcomes such as watch time while searching large item corpora efficiently.
We present the Controllable Retrieval Model (CRM), an outcome-conditioned retrieval framework that adds watch time as an input to the user tower.
During training, CRM learns to predict the next item given its observed watch time; at inference, we supply a desired watch time to steer retrieval without retraining.
The item tower is independent of the condition, so item embeddings can still be precomputed and indexed for approximate nearest neighbor (ANN) search.
We implement CRM as both a DNN-based two-tower model and a Decision Transformer.
The Transformer interleaves item tokens with watch-time-to-go condition tokens, using the target watch time as a numerical prompt for next-item prediction.
For each online request, we choose between the average and maximum watch times in the user's recent history as the condition.
CRM is deployed in Kuaishou's short-video recommendation system, which serves over 400 million users.
Online A/B tests show video watch time improvements of 0.372\% in Kuaishou and 0.457\% in Kuaishou Lite.
\end{abstract}

\begin{CCSXML}
<ccs2012>
<concept>
<concept_id>10002951.10003317.10003347.10003350</concept_id>
<concept_desc>Information systems~Recommender systems</concept_desc>
<concept_significance>500</concept_significance>
</concept>
</ccs2012>
\end{CCSXML}

\ccsdesc[500]{Information systems~Recommender systems}

\keywords{Outcome-conditioned retrieval, condition tokens, generative recommendation, prompt-based recommendation, short-video recommendation}

\maketitle
\pagestyle{plain}
\thispagestyle{plain}
\raggedbottom
\hypersetup{
  pdfauthor={Chi Liu, Jiangxia Cao, Rui Huang, Kuo Cai, Weifeng Ding, Qiang Luo, Kun Gai, Guorui Zhou},
  pdfkeywords={Outcome-conditioned retrieval, condition tokens, generative recommendation, prompt-based recommendation, short-video recommendation}
}

\begin{figure*}[t]
  \centering
  \includegraphics[width=\textwidth]{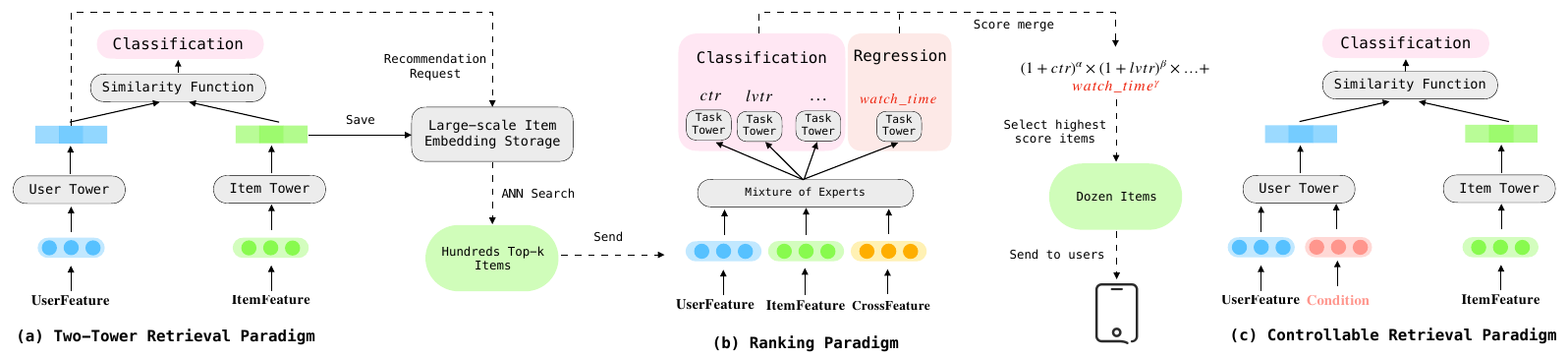}
  \caption{Retrieval and ranking in a recommendation system: (a) a standard two-tower retrieval model trained with a classification objective; (b) a ranking model with classification and regression objectives; (c) CRM, which adds an outcome condition to the user tower.}
  \Description{Three architecture diagrams compare standard two-tower retrieval, multi-task ranking, and CRM. CRM adds a condition to the user tower and retains an independent item tower and ANN search.}
  \label{intro}
\end{figure*}

\section{Introduction}

Recommendation systems serve billions of users on platforms such as Kuaishou~\cite{kuaiformer}, Douyin~\cite{clock}, and Google~\cite{tiger}.
Industrial systems optimize several objectives~\cite{ple, home}, including discrete feedback such as clicks and comments, and continuous outcomes such as watch time.
They must balance these objectives when selecting items from a corpus that can contain tens of millions of candidates.

A typical system has two stages: retrieval~\cite{ebr} selects hundreds of candidates, and ranking~\cite{youtubednn} scores them against multiple objectives to select the items shown to the user (Figure~\ref{intro}(a) and (b)).
Retrieval often uses a two-tower model~\cite{dssm, yang2020mixed}: separate user and item encoders are trained with a classification objective.
Item embeddings can be computed in advance, so each request requires only a new user embedding and a nearest neighbor search.
To support multiple objectives, systems commonly use several retrieval routes, each trained for a different target, and merge their candidates before ranking~\cite{huang2024comprehensive}.

Continuous outcomes pose a different challenge.
Expected watch time depends on both the user and the video, whereas the user tower computes its representation before any candidate is selected.
There is also no single watch-time threshold that captures a good outcome across users and videos.
This makes it difficult to translate the ranking stage's watch-time objective into a retrieval query.

We propose the Controllable Retrieval Model (CRM), an outcome-conditioned retrieval framework for recommendation (Figure~\ref{intro}(c)).
CRM uses watch time as an input to the user tower: training uses the observed watch time of the target video, while inference uses a desired watch time to steer retrieval.
The condition can be changed without retraining the model.
Item embeddings remain independent of the condition and can still be indexed for approximate nearest neighbor (ANN) search.
We implement CRM with a DNN-based two-tower model and a Decision Transformer-based model~\cite{dt} that interleaves watch-time-to-go condition tokens with item tokens.
Like generative recommendation models, the Transformer predicts the next item from the interaction sequence, with a numerical watch-time prompt specifying the desired outcome.

Our contributions are as follows:
\begin{itemize}[leftmargin=*,align=left]
    \item We introduce watch-time conditioning into two-tower retrieval, allowing the target to be set at inference while retaining precomputed item embeddings and ANN search.
    \item We implement CRM with a DNN and a Transformer that interleaves watch-time-to-go condition tokens with item tokens.
    \item We combine average and maximum watch-time conditions in an online serving strategy and evaluate CRM in Kuaishou's largest short-video recommendation scenario.
\end{itemize}

\section{Preliminaries}

\subsection{Two-Tower Retrieval Workflow}
The retrieval task is to predict which video a user will interact with next, using user and item features.
Let ${(x_1, w_1), \dots, (x_n, w_n)}$ denote the user's recent history, where $x_i$ is a video ID and $w_i$ is its observed watch time.
This history forms part of \texttt{UserFea}, and the target video $x_{n+1}$ is represented in \texttt{ItemFea}.
The two towers encode these features separately:
\begin{equation}
\begin{split}
 \mathbf{u} &= \texttt{User\_Tower}\big(\texttt{UserFea}\big),\\
 \mathbf{v} &= \texttt{Item\_Tower}\big(\texttt{ItemFea}\big),\\
\end{split}
\label{twotower}
\end{equation}
where $\texttt{User\_Tower}$ and $\texttt{Item\_Tower}$ can be networks such as multilayer perceptrons (MLPs).
Using long-view prediction as an example, we minimize:
\begin{equation}
\begin{split}
&\mathcal{L} = -\texttt{log}{(\hat{y}^{\texttt{lvtr}})}, \\
\texttt{where}\quad \hat{y}^{\texttt{lvtr}} &= \texttt{In-batch-Softmax}(\mathbf{v}^\top\mathbf{u}),
\end{split}
\label{loss}
\end{equation}
where $\hat{y}^{\texttt{lvtr}}$ is the softmax probability assigned to the target item among the in-batch candidates.

At inference, we cache the item embeddings in $\mathbf{V}$ and compute the user embedding $\mathbf{u}$ for each request.
We then retrieve candidates by nearest neighbor search:
\begin{equation}
 \{\hat{x}_{n+1}^1, \hat{x}_{n+1}^2, \hat{x}_{n+1}^3, \dots\} = \texttt{ANN}(\mathbf{u}, \mathbf{V}),
\label{infer}
\end{equation}
where \texttt{ANN} is an approximate nearest neighbor search operator~\cite{faiss}, and $\hat{x}_{n+1}^i$ denotes a retrieved candidate.

\subsection{Decision Transformer in RL}
Reinforcement learning (RL) learns a policy that selects actions to maximize cumulative reward~\cite{levine2020offline, mnih2013playing}.
In recommendation, the user's interaction history can represent the state, the recommended item is an action, and user feedback supplies the reward.
The Decision Transformer (DT)~\cite{dt} formulates offline RL as sequence modeling over logged trajectories.
Its input interleaves the return-to-go $R_i$, state representation $\mathbf{s}_i$, and action $a_i$, as shown in Figure~\ref{decision}(a):
\begin{equation}
\begin{split}
 a_{n+1} &= \texttt{DT}\big(R_1, \mathbf{s}_1, a_1, R_2, \mathbf{s}_2, a_2,\dots, R_{n+1}, \mathbf{s}_{n+1} \big),\\
\end{split}
\label{eq:decision}
\end{equation}
where $\texttt{DT}$ is a multilayer Transformer with a causal attention mask and $a_{n+1}$ is the predicted next action.
For a trajectory ending at step $n+1$, the return-to-go is $R_i=\sum_{j=i}^{n+1}r_j$, where $r_j$ is the reward at step $j$.
Conditioning on the return-to-go allows the model to predict actions for a specified reward target.

\section{Methodology}

\subsection{Offline Training Strategy}
CRM incorporates watch-time conditions into either a DNN-based user tower or a Transformer over the user's interaction history.

\subsubsection{DNN-based CRM}
The DNN-based model adds the observed watch time of the next video, $w_{n+1}$, to the user tower's inputs (Figure~\ref{intro}(c)).
It learns to predict the target item conditioned on the user's features and this watch time:
\begin{equation}
\begin{split}
 \mathbf{u}^{\texttt{Condition}} &= \texttt{CRM\_User\_Tower}\big(\texttt{UserFea}_u, w_{n+1}\big),\\
\end{split}
\label{crmdnn}
\end{equation}
where $\mathbf{u}^{\texttt{Condition}}$ is the resulting user embedding.
We train CRM with the two-tower loss in Eq.~(\ref{loss}).
The item tower and retrieval procedure are unchanged.

\begin{figure}[t]
  \centering
  \includegraphics[width=8cm,height=7cm]{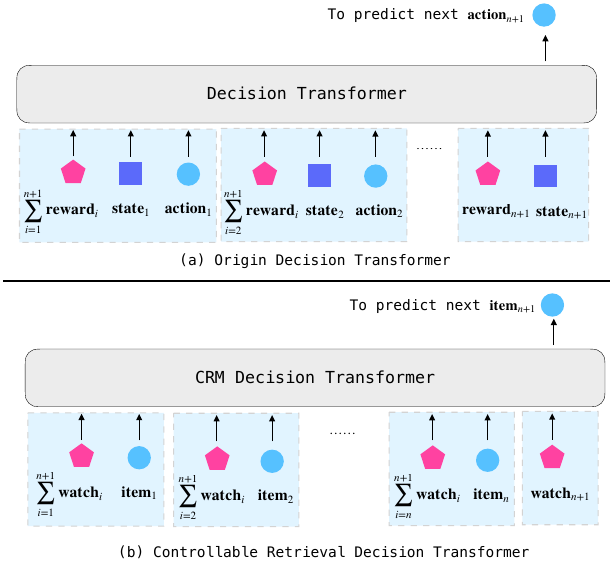}
  \caption{Input sequences for (a) the original Decision Transformer and (b) CRM. The original model uses return-to-go, state, and action tokens. CRM alternates item tokens with watch-time-to-go condition tokens. Both models use causal attention masks.}
  \Description{The original Decision Transformer interleaves return-to-go, state, and action tokens. CRM alternates watch-time-to-go and item tokens to predict the next item.}
  \label{decision}
\end{figure}

\subsubsection{Transformer with Condition Tokens}
We also use sequence modeling~\cite{sutskever2014sequence, achiam2023gpt} to predict the next item from the user's interaction history.
Following the Decision Transformer~\cite{dt}, we condition this prediction on cumulative watch time (Figure~\ref{decision}(b)).
The input alternates watch-time-to-go condition tokens $W_i$ and item tokens $x_i$, ending with the condition for the next item:
\begin{equation}
\begin{split}
\mathbf{u}^{\texttt{Condition}} = \texttt{CRM\_User}&\texttt{\_Tower}\big(\texttt{UserFea}_u, \\
&\texttt{CRM\_DT}(W_1, x_1, W_2, x_2,\dots, W_{n+1})\big),\\
\end{split}
\label{crmtransformer}
\end{equation}
where $\texttt{CRM\_DT}$ is a multilayer Transformer with a causal attention mask, and $W_i=\sum_{j=i}^{n+1}w_j$ is the cumulative watch time from step $i$ to step $n+1$.
During training, we compute $W_i$ from observed watch times.
At inference, we set $W_{n+1}$ to the desired watch time for the next item.
We use the Transformer version in production.

\begin{algorithm}[t]
\caption{Online condition selection.}
\label{algo:online}
\begin{algorithmic}[1]
\REQUIRE Watch history $\{(x_1, w_1),\dots, (x_n, w_n)\}$, length $n$, probability $p$.
\ENSURE Watch-time condition.
\STATE SumValue = 0,\  MaxValue = 0
\FOR {$i$ from $1 \rightarrow {n}$}
\STATE SumValue = SumValue + $w_i$
\STATE MaxValue = \texttt{max}(MaxValue, $w_i$)
\ENDFOR
\STATE AvgValue = SumValue / $n$
\IF{\texttt{RandomFloat} < $p$}
\RETURN MaxValue
\ENDIF
\RETURN AvgValue
\end{algorithmic}
\end{algorithm}

\subsection{Online Inference Strategy}
At inference, we choose the target watch time from the user's recent history, using either the average or the maximum over the last $n$ videos.

\subsubsection{Maximum vs. Average Watch Time}
The average watch time reflects the user's recent viewing behavior, while the maximum sets a higher target to encourage longer viewing.

\begin{figure}[t]
  \centering
  \includegraphics[width=8cm,height=4cm]{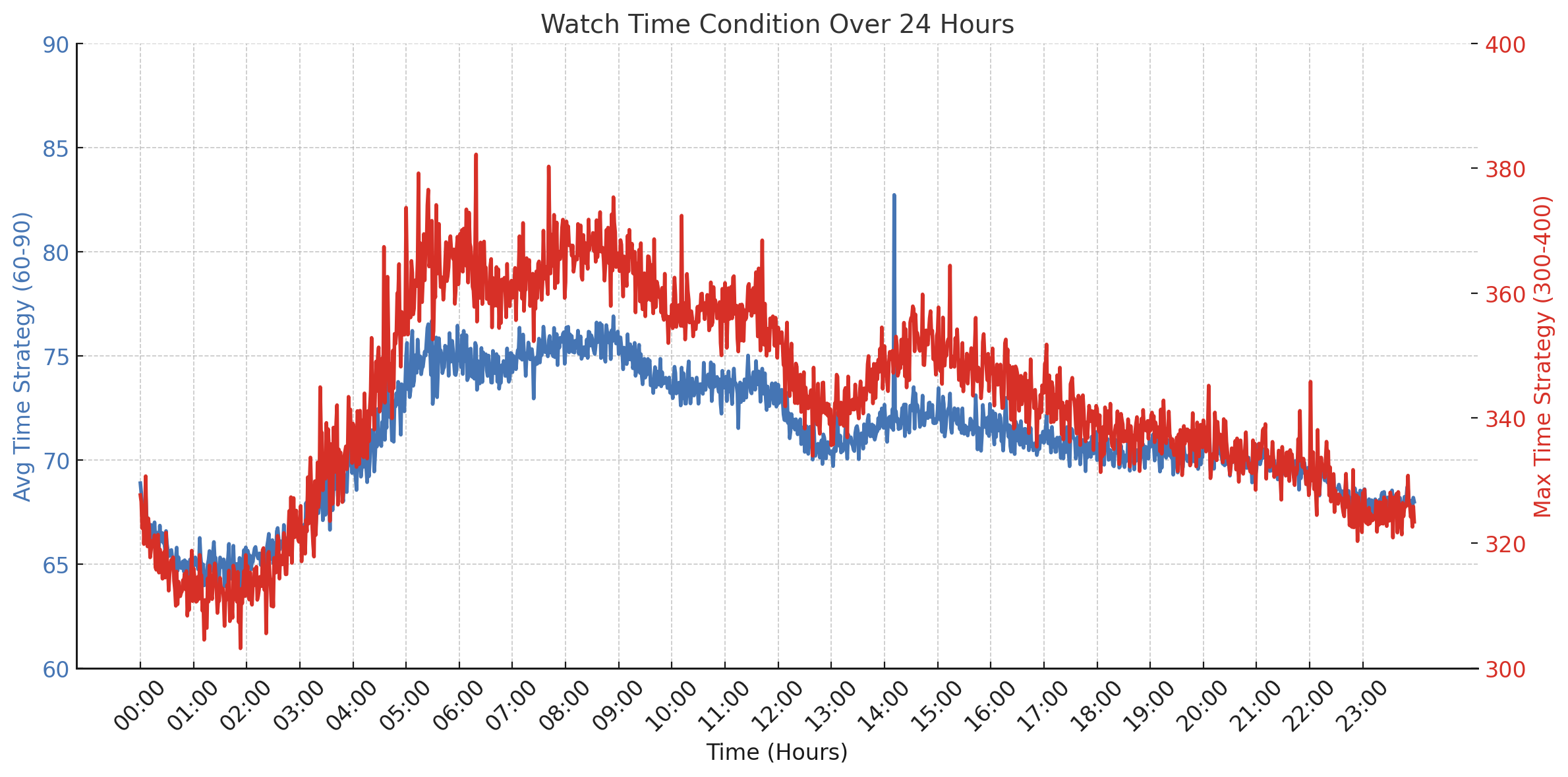}
  \caption{Average and maximum watch-time conditions over a day.}
  \Description{Two curves show the average and maximum watch-time conditions over 24 hours. They follow similar daily trends, with the maximum condition higher than the average.}
  \label{tab:condition_variation}
\end{figure}

The history length $n$ affects both choices.
A large window can make the maximum sensitive to an old outlier, while a small window may not represent the user's viewing habits.
We use $n=32$ in production.
Figure~\ref{tab:condition_variation} shows that the average and maximum conditions follow similar daily trends, with the maximum consistently higher.

\begin{table*}[ht]
\centering
\setlength{\tabcolsep}{6pt}
\begin{tabular}{ccccccccc}
\toprule
\multirow{2}{*}{Applications} & \multirow{2}{*}{\makecell{Video \\Watch Time}} & \multirow{2}{*}{\makecell{Total App\\Usage Time}} & \multirow{2}{*}{\makecell{Usage Time\\per User}} & \multirow{2}{*}{\makecell{Avg. Time per\\Video View}} & \multirow{2}{*}{\makecell{Video\\Views}} & \multirow{2}{*}{\makecell{Likes}} & \multirow{2}{*}{\makecell{Follows}} & \multirow{2}{*}{\makecell{Comments}} \\
\\
\midrule
\makecell{Kuaishou\\Single Page}& +0.372\% & +0.175\% & +0.144\% & +0.323\% & +0.048\% & +0.029\% & +0.447\% & +0.330\% \\
\midrule
\makecell{Kuaishou Lite\\Single Page} & +0.457\% & +0.196\% & +0.122\% & +0.133\% & +0.324\% & +0.137\% & +0.828\% & +0.697\% \\
\bottomrule
\end{tabular}
\caption{Online A/B test results for short-video recommendation at Kuaishou.}
\label{mainonline}
\end{table*}

\subsubsection{Time-Division Multiplexing Strategy}
In our online experiments, the average condition increased video views and interaction metrics such as comments and likes, but brought little improvement in total watch time.
The maximum condition increased watch time, but tended to retrieve longer videos and could reduce video views and interactions.

We combine the two through time-division multiplexing: for each request, we choose the maximum condition with probability $p$ and the average condition with probability $1-p$ (Algorithm~\ref{algo:online}).
We select $n$ and $p$ through online experiments.

\section{Experiments}

We evaluate CRM in Kuaishou's largest short-video recommendation scenario, which serves over 400 million users and generates 50 billion logs per day.
We address two questions:
\begin{itemize}[leftmargin=*,align=left]
    \item \textbf{RQ1}: How does CRM affect online recommendation metrics?
    \item \textbf{RQ2}: How does CRM compare with other retrieval methods?
\end{itemize}

\begin{figure}[t]
  \centering
  \includegraphics[width=8cm,height=4.5cm]{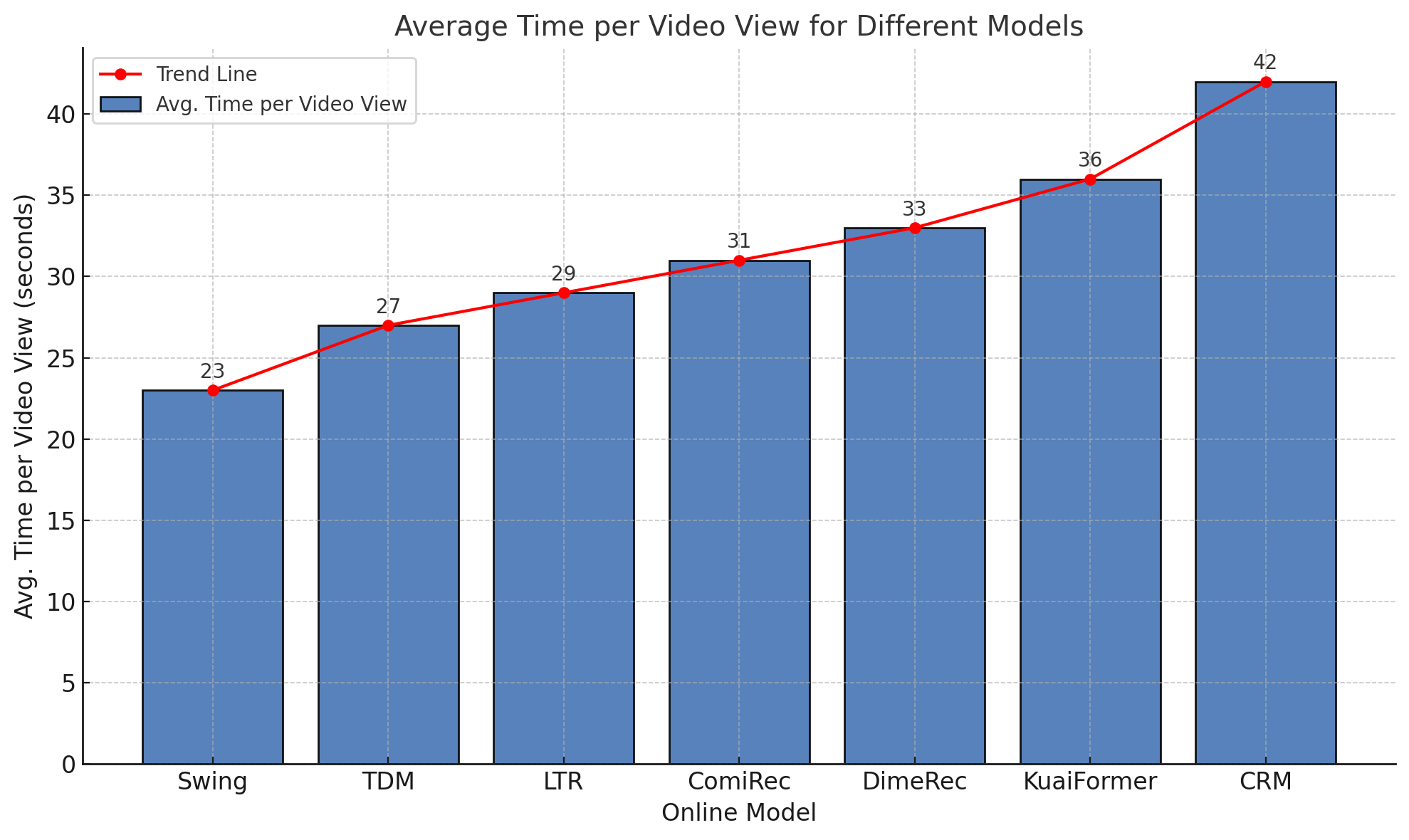}
  \caption{Average watch time per video view for different retrieval models.}
  \Description{Average watch time in seconds: Swing 23, TDM 27, LTR 29, ComiRec 31, DimeRec 33, KuaiFormer 36, and CRM 42.}
  \label{tab:avg_time_per_video_view_sorted}
\end{figure}

\subsection{Online A/B Test (RQ1)}
Table~\ref{mainonline} reports the online A/B test results.
CRM increased video watch time by 0.372\% in Kuaishou and 0.457\% in Kuaishou Lite.
Total app usage time increased by 0.175\% and 0.196\%, respectively, with gains in usage time per user in both applications.

Video views, likes, follows, and comments also increased in both applications.
In Kuaishou Lite, video views increased by 0.324\% and follows by 0.828\%.

\subsection{Comparison with Other Retrieval Methods (RQ2)}
We compare CRM with six retrieval methods deployed in the same short-video recommendation scenario:
(1) Swing~\cite{swing}, an item-to-item method;
(2) ComiRec~\cite{comirec}, a multi-interest model;
(3) DimeRec~\cite{dimerec}, a diffusion model;
(4) LTR~\cite{zheng2024full}, a list-wise method;
(5) TDM~\cite{tdm}, a tree-based retrieval method; and
(6) KuaiFormer~\cite{kuaiformer}, a Transformer-based model.

Figure~\ref{tab:avg_time_per_video_view_sorted} compares average watch time per video view.
CRM reaches 42 seconds, the highest value among the compared methods.

\section{Conclusion}
CRM makes the desired watch time an input to retrieval while retaining precomputed item embeddings and ANN search.
We implement this through a watch-time feature in the DNN model and condition tokens in the Transformer.
Online experiments at Kuaishou show gains in watch time, video views, and interactions.
Future work will explore other outcome conditions and how to choose them for each request.

\balance
\bibliographystyle{ACM-Reference-Format}
\bibliography{references}
\end{document}